\documentclass[11pt]{article}

\usepackage[pdftex]{graphicx}
\DeclareGraphicsExtensions{.jpg,.jpeg,.png}

\usepackage{listings}
\usepackage{color}
\usepackage[dvipsnames]{xcolor}
\usepackage{algorithmic}
\usepackage{breqn}
\usepackage{verbatim}
\usepackage{hyperref}
\usepackage{authblk}

\begin{document}

\title{A Security Analysis and Revised Security Extension for the Precision Time Protocol}

\author{Eyal Itkin}
\author{Avishai Wool}

\affil{{\tt eyalitki@post.tau.ac.il}, {\tt yash@eng.tau.ac.il}\\Tel Aviv University, Israel}

\renewcommand\Affilfont{\itshape\small}

\maketitle

\begin{abstract}

The Precision Time Protocol (PTP) aims to provide highly accurate and synchronized clocks. Its defining standard, IEEE 1588, has a security section (``Annex K'') which relies on symmetric-key secrecy. In this paper we present a detailed threat analysis of the PTP standard, in which we highlight the security properties that should be addressed by any security extension. During this analysis we identify a sequence of new attacks and non-cryptographic network-based defenses that mitigate them. We then suggest to replace Annex K's symmetric cryptography by an efficient elliptic-curve Public-Key signatures. We implemented all our attacks to demonstrate their effectiveness, and also implemented and evaluated both the network and cryptographic defenses. Our results show that the proposed schemes are extremely practical, and much more secure than previous suggestions.

\end{abstract}


\section{Introduction}\label{Introduction}

\subsection{PTP Overview}\label{PTP}

IEEE 1588 is a standard defining the Precision Time Protocol (PTP), \cite{standard}. An initial version of the standard was published at 2002, and a second version was later published in 2008. PTP is aimed specifically at measurement, control and financial applications, applications that have an increasing need for highly accurate and synchronized clocks. The protocol was built to address the specific requirements of these types of applications:

\begin{enumerate}

  \item Spatially localized - A factory computer network for instance

  \item Microsecond to sub-microsecond accuracy and precision

  \item Administration free - plug-and-play distributed algorithm

  \item Accessible for both high-end devices (servers) and low-cost, low-end devices (basic controllers)

\end{enumerate}

In contrast to the centralized Network Time Protocol (NTP) \cite{rfc958}, that only reaches an accuracy level of several milliseconds, the PTP standard defines a distributed network of clocks, called \emph{PTP nodes}, that dynamically builds a master-slave hierarchy, which collectively achieves the desired accuracy. However, PTP has some similarities to NTP: both protocols allow each node to calculate the unique network delay between itself and the central clock. This central clock is called the \emph{Grand Master}.

One of the main novelties of PTP is the fact that the design has provisions for the use of hardware timestamping, as explained in \cite{version2}. From the need to reach very high accuracy and precision, rose the need to bypass the jitter, or noise, induced by the message passing through the network stack in the OS. This is done using PTP-aware network interface cards (NICs), that are able to timestamp the message in the lowest hardware levels, just before the message is sent to the underlying physical layer. Note that hardware timestamping breaks most cryptographic primitives unless special steps are taken.

\subsection{Related Work}\label{related}

Since the publication of Annex K, there were several calls for improvements and even calls to adopt a brand new security extension for the protocol. Most of the papers regarding the security of the protocol where published by the members of the IEEE 1588 security subcommittee, as summarized in \cite{state_of_the_art}. In 2011, Mizrahi \cite{ISPCS_2011} suggested to make use of the known IPSec solution, and to base the security Type-Length-Value (TLV) extension header on it. Later on, in 2012, after the results of \cite{ISPCS_2012}, Kirrmann suggested a modified Annex K. His main changes were to remove the redundant 3 way-handshake, to update the Message Authentication Code (MAC) that is used in the Integrity Check Value (ICV) calculation and to use IPSec for the key distribution scheme. A proposal by Ellegaard \cite{Ellegaard}, speaks about hop-by-hop security based on the MACsec. These proposals speak only about the means to establish a secure transport channel between the 2 PTP nodes, regardless of the nodes' role in the protocol, or the protocol's behavior.

A proposal by Fries \cite{Fries}, speaks mainly on the key distribution scheme, and suggests to use a standard cryptographic algorithms for the ICV calculation. Sibold/Dickerson \cite{Sibold} \cite{Dickerson} suggest using TESLA \cite{TESLA} for the Security Association (SA) and to check the applications of Network Time Security (NTS), a standard that is currently underway. While NTS presents a new asymmetric approach, master X509 certificates, the main use of it is for the bootstrapping of TESLA.

As we can see, the papers mainly address the key distribution problem, that is not part of the original Annex K, or suggest new ways to establish a secure channel between the PTP nodes. We argue that one of the main design flaws in Annex K is its inability to handle an insider threat, as will be shown in the next sections, and the related works do not address this issue.

Nevertheless, in \cite{state_of_the_art} there is a suggestion for a new methodology, unlike \cite{rfc7384}, based on a ``divide and conquer'' technique. In an internal paper of the security subcommittee, \cite{Prongs}, they present a ``four pronged approach'' which suggests 4 level of security:

\begin{itemize}

  \item Prong A - End-to-End strong source authentication, based on a TLV for the PTP standard.

  \item Prong B - hop-by-hop security mechanism, external to PTP.

  \item Prong C - Architectual guidance defining the redundant communication links, to handle selective delay attacks.

  \item Prong D - Monitoring and management guidance.

\end{itemize}

We chose to follow the ``four pronged approach'', and present a fully functioning \emph{Prong A} security extension for IEEE 1588.

\subsection{Contributions}\label{contributions}


%

In this paper we present a detailed threat analysis of the PTP standard, in which we highlight the security properties that should be addressed by any security extension. We identified a sequence of new attacks that can be mounted against PTP, depending on the attacker's strength: from out-of-band network attacker to corrupt insiders. We then identified network defenses that mitigate some of the attacks: binding the clock-IDs to the network addresses, and expanded use of message sequence numbers to prevent spoofing and introduce session-like semantics into the protocol. While the basic version of these defenses is fully compatible with the PTP message format, we suggest a small modification - to use 2 reserved bytes and extend the sequence ID to 32 bits - which greatly improves the protocol's resilience to attack.

We then suggest to replace Annex K's symmetric cryptography by an efficient elliptic-curve Public-Key signature scheme. The proposed EdDSA signatures are only 64 bytes long, and only need to be attached to some of the grandmaster messages. We implemented all our attacks to demonstrate their effectiveness, and also implemented and evaluated both the network and cryptographic defenses. Our results show that the proposed schemes are extremely practical, and much more secure than previous suggestions.

\textbf{Organization} In the next section we introduce the main aspects of PTP. Section 3 presents the adversaries that play a role in our threat analysis, while section 4 presents all the relevant attack scenarios and completes the threat analysis. In Section 5 there are full details of our proposed security extension. We then show experimental results of our defense mechanism, and compare it to the state of the art security proposals. We conclude with Section 6.


\section{Preliminaries}

The protocol consists of two main layers, and an additional management layer:

\begin{enumerate}

  \item Grandmaster election - the Best Master Clock (BMC) algorithm

  \item Time oriented messages - Sending timestamps and measuring network delay

\end{enumerate}

The first layer is a distributed leader election algorithm that is being constantly calculated by all of the master-candidate PTP nodes. Each candidate node sends an \texttt{ANNOUNCE} message in which it declares its management priority, its time source, etc., and the best clock is elected as the ``grandmaster'' clock. The leader election algorithm is called ``Best Master Clock'' algorithm. The elected master clock then multicasts its accurate timestamp to all of the slave nodes, using the 2nd layer of the protocol.

The 2nd layer of the protocol consists of the following messages:

\begin{enumerate}

  \item \texttt{SYNC} - The master's timestamp, broadcast to all PTP nodes every \#seconds

  \item \texttt{DELAY_REQ} - A call from a slave to measure the delay with the master

  \item \texttt{DELAY_RESP} - An answer from the master with its receive timestamp

\end{enumerate}

\begin{figure}[t]

\centerline{\includegraphics[width=4.5in]{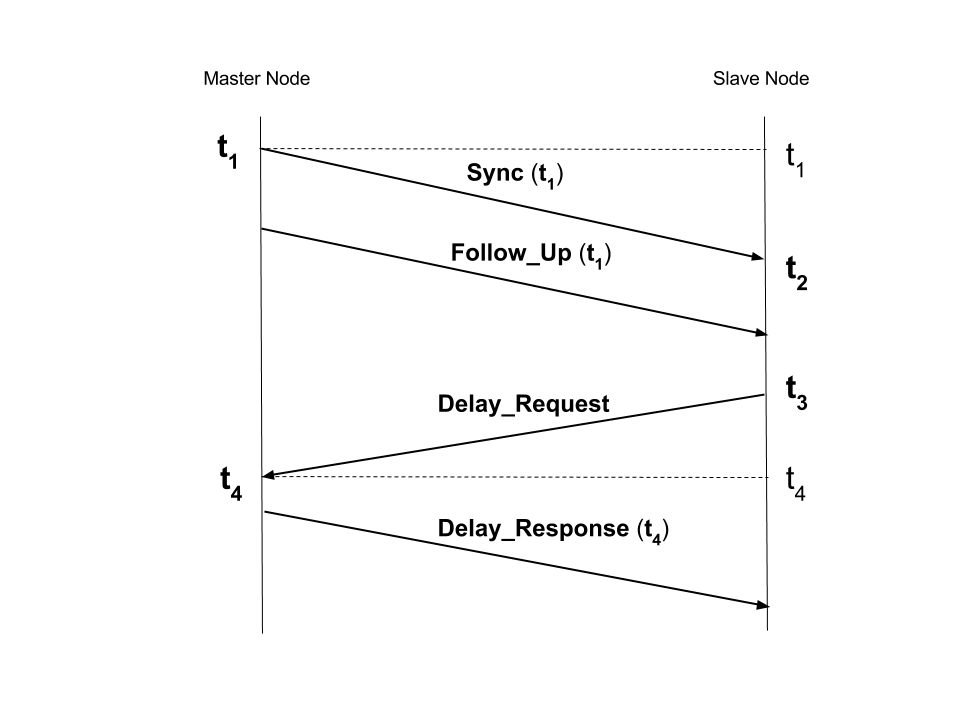}}

\caption{PTP 2-step message flow. Note the 4 timestamps $t_1$, $t_2$, $t_3$, $t_4$: the \texttt{FOLLOW_UP} and \texttt{DELAY_RESP} messages carry $t_1$ and $t_4$ to the slave.}

\label{Messages}

\end{figure}

As can be seen in Figure \ref{Messages}, the master periodically sends \texttt{SYNC} messages with its timestamp ($t_1$), and the slave receives them ($t_2$). In addition, once in a while the slave measures its delay from the master by sending a \texttt{DELAY_REQ} ($t_3$), and the master replies with \texttt{DELAY_RESP} including its receive timestamp ($t_4$). Using these 4 time points, the slave can calculate the delay between the master and itself:

\begin{equation}
  delay = \frac{(t_2 - t_1) + (t_4 - t_3)}{2}
\end{equation}

And the offset in clock rates between the two clocks:

\begin{equation}
  offset = \frac{(t_2 - t_1) - (t_4 - t_3)}{2}
\end{equation}

Clocks that agree on their delay, or latency, are called \emph{synchronized}, and clocks that agree on their clock rates are called \emph{syntonized}. Deducing the offset in clock rates is done by slave during the protocol. This will be used for multiplying the slave's clock ticks as a compensation for the offset in clock rates.

\subsection{2-Step sync}\label{2step}

Hardware timestamping mechanism breaks support for most cryptographic primitives since encryption or signatures applied by a higher level in the software stack must happen \textbf{before} the message is sent. E.g., if the NIC hardware modifies a signed message in transit - the receiver will be unable to verify the signature. Therefore, the standard defines a notion of \emph{2-step} communication. In such cases, the \texttt{SYNC} message is just a placeholder for the $t_1$ time point. A PTP-aware NIC recognizes the \texttt{SYNC} message, calculates $t_1$ and stores it locally. Then the software layers query the NIC, obtain $t_1$, and embed its value in the body of the \texttt{FOLLOW_UP} message that will immediately follow, see Figure \ref{Messages}. This solution enables cryptographic operations to be done on the timestamp that is now known to the grandmaster's software stack.

\subsection{PTP Annex K}\label{appendix}

IEEE standard 1588 defines an experimental security extension in its Annex K. The goals of the security protocol are: providing group source authentication, message integrity, and prevention of replay attacks. According to these goals the protocol was designed to consist mainly of two parts:

\begin{enumerate}

  \item A Challenge-Response mechanism used to affirm the trust relations between PTP nodes

  \item An HMAC \cite{HMAC} based integrity protection code, using a shared symmetric key, using anti-replay counters

\end{enumerate}

The Annex K security protocol consists of the following steps. First, each legitimate PTP node in the network starts with a preshared symmetric key, (the key distribution scheme is out of the scope of Annex K). Second, the communication itself is done between unidirectional Security Associations (SA). Each SA consists of source port + network address, destination, key, random lifetime (freshness) and a replay counter. Third, when a node sends messages to an untrusted destination node, it first performs a 3-way handshake in which both sides authenticate to each other using the preshared key. After the challenge-response message transfer finishes, both nodes have established trust relationships and have matching SAs. Fourth, every message sent over to a given SA has an authentication Type-Length-Value (TLV) extension header, that contains an Integrity Check Value (ICV) that is calculated over all of the message, including anti-replay counters. Fifth, after a predetermined amount of time the SA becomes ``old'' and the nodes needs to ``freshen'' the trust relations. This step ensures that the replay counters won't roll over.

An important note is that the Annex's goals are only defined against non-PTP computers on the network. By its symmetric nature, using ``group source authentication'', the security protocol has no effective way to differentiate between PTP clocks: any PTP slave can easily masquarade as the grandmaster. The protocol only defines means to build a cryptographicaly secure unicast channel between 2 PTP nodes, and does not address any attempt of fraud over the established channel.

Although the second version of the standard was published in 2008, Annex K was never really adopted by the industry, as was mentioned in \cite{first_deploy}, and it has quite a few drawbacks, as specified in \cite{state_of_the_art}. Due to its deployment status, the initial experimental status and the public debates about its efficiency and effectiveness, Annex K does not seem like a promising security extension for the PTP standard.

\subsection{Testing Environment}\label{environment}

In our work we used \emph{DeterProject}'s infrastructure, \cite{Deter} and \cite{DeterProject}, in order to experiment and test our results. The experimentation was done on the open source linux PTP daemon, \emph{PTPd} \cite{PTPd}. We implemented our proposed security extension on top of the daemon's 2.3.1.1 version, in addition to the creation of a python scapy layer supporting the PTP protocol, including our extension.


\section{Precision Time Protocol Threat Analysis: Adversaries}\label{Threat Analysis}

In this section we discuss the different adversaries that can harm the protocol. The adversaries are listed in an increasing order of strength, from the weakest \emph{Out-of-Band applicative adversary} to the most powerful \emph{hostile grandmaster clock adversary}. We separate the attackers into 2 categories: \emph{Outsiders} and \emph{Insiders}.

\subsection{Out-of-Band Applicative adversary}\label{adv:OOB}

The Out-of-Band (OOB) Applicative adversary is located out of the path of the protocol's unicast messages. The adversary can only see the public multicast protocol messages: \texttt{ANNOUNCE} and \texttt{SYNC}. This weak adversary can only forge fake PTP layers of messages, and it is not capable of faking lower-level networking layers, such as UDP/Ethernet.

Although this is the weakest available adversary, it is worth mentioning that any attack that it performs will most likely not be intercepted by any IDS/IPS deployed on the network, unless it has PTP-aware capabilities.

\subsection{Out-of-Band Network adversary}\label{adv:OOB-net}

The Out-of-Band (OOB) Network adversary is a stronger version of the previous adversary. This adversary now has the capability to also forge fake network layers, i.e., spoof IP and Ethernet headers. Note that this is still a very weak adversary since Layer 2 and Layer 3 spoofing is quite trivial to preform.

\subsection{In-Band adversary}\label{adv:IB}

The In-Band (IB) adversary is located in the path of all of the protocol's messages, both unicast and multicast. This Man-In-The-Middle (MITM) position between the targeted PTP slave node and its associated master gives the attacker full control over the communication channel: dropping messages, inserting fake messages and changing fields of passing messages.

An adversary can gain such a strong position, for instance, by:

\begin{enumerate}

  \item Deploying low-level network attacks, such as ARP-poisoning or DNS-poisoning, in order to ``bend'' the traffic toward itself

  \item Targeting a middle box to gain control over a switch, router or public hotspot

\end{enumerate}

This is the strongest attacker that is not a legitimate part of the PTP network.

\subsection{Hostile PTP Insider clock}\label{adv:slave}

The Hostile PTP Insider clock adversary is a legitimate PTP slave node on the network, that is controlled by a hostile adversary. This clock has poor clock parameters, thus it can't win the BMC leader election in a legitimate way.

An adversary can gain such a position more easily than can be suspected: PTP aims to accurately synchronize the time between a large number of computers and machines in a local communication network. Hence, every such computer is a legitimate PTP node, and now a possible entrance point to the logical PTP network. As a result, gaining access to the PTP network can be done by an insider threat in the original network, or by an adversary that launches a dedicated, non-PTP, attack on one of the network participants. This is the weakest insider attacker.

\subsection{Hostile PTP Management node}\label{adv:tech}

The Hostile PTP management node adversary is a legitimate PTP management node on the network, that is controlled by a hostile adversary. Such an adversary holds all the protocol's management capabilities, resulting in a very powerful adversary. In this work we won't address the threats resulting from such a strong adversary. We suggest to deploy standard security and monitoring tools to ensure the legitimacy of the management node in the network.

\subsection{Hostile PTP Grandmaster node}\label{adv:master}

The Hostile PTP grandmaster node adversary is a legitimate PTP clock node on the network, that is controlled by a hostile adversary. In contrast to the Hostile PTP Insider clock, this adversary now controls the clock with the best PTP parameters in the network, meaning that the clock can legitimately win the BMC leader election. Such an adversary holds all of the protocol's time distribution capabilities, resulting in an all powerful adversary. In this work we won't address the threats resulting from such a strong adversary. We suggest to deploy security-hardened PTP middle boxes, in addition to deploying external monitoring tools, in order to ensure the legitimacy of the grandmaster PTP node in the network.


\section{Attack Scenarios and Mitigations}\label{Attack Scenarios}

In the following subsections we discuss the results of our analysis. The analysis describes a continuous arms race between the system's adversaries and our suggested mitigations against those attacks.

The Precision Time Protocol consists of 3 main protocol layers:

\begin{enumerate}

  \item Management Layer

  \item Event messages - Time oriented Messages

  \item Best Master Clock Algorithm - grandmaster leader election

\end{enumerate}

In the following subsections we introduce the 3 main arms races between the protocol's adversaries and the proposed defense mitigations. Each arms race corresponds to a different protocol layer, thus covering all of the protocol's main layers.


\subsection{Event Messages Attacks}\label{attack:time}

The IEEE 1588 standard separates the messages into two main groups:

\begin{enumerate}

  \item Event Messages, i.e., time oriented messages, such as \texttt{SYNC}

  \item General Messages, i.e., management orient messages, such as \texttt{ANNOUNCE}

\end{enumerate}

In this subsection we focus on the attack scenarios that target the time synchronization of the PTP nodes. These types of attacks have received the majority of security-aware attention so far, as in \cite{Threat_Analysis} and \cite{Delay}, because of the dominant role of these messages in the overall protocol. In the following arms race we discuss our suggested defense mitigations, and show the experimental results of their deployment.

\subsubsection{Delay Spoofing Attack Scenario}\label{time:scenrio1}

\paragraph{Attack}

Forge \texttt{DELAY_RESPONSE} messages on behalf of the grandmaster, thus damaging the delay measurements of the target node. Can be performed by the OOB Applicative adversary, or any other stronger adversary.

Note that a \texttt{DELAY_RESPONSE} message has a \texttt{sequenceId} field - so the OOB adversary needs to spoof messages with a matching \texttt{DELAY_REQUEST} \texttt{sequenceId} field. Hence, first he needs to bend the unicast \texttt{DELAY_REQUEST} towards himself. This can be done by replaying the grandmaster's multicast messages, \texttt{SYNC} or \texttt{ANNOUNCE}, using its own network address. Such a replay results in the attacker's address being registered at the slave as the matching network identity of the master. In the PTPd implementation this can easily be done using the master's \texttt{ANNOUNCE} messages.

\begin{figure}[t]

\centerline{\includegraphics[width=4.5in]{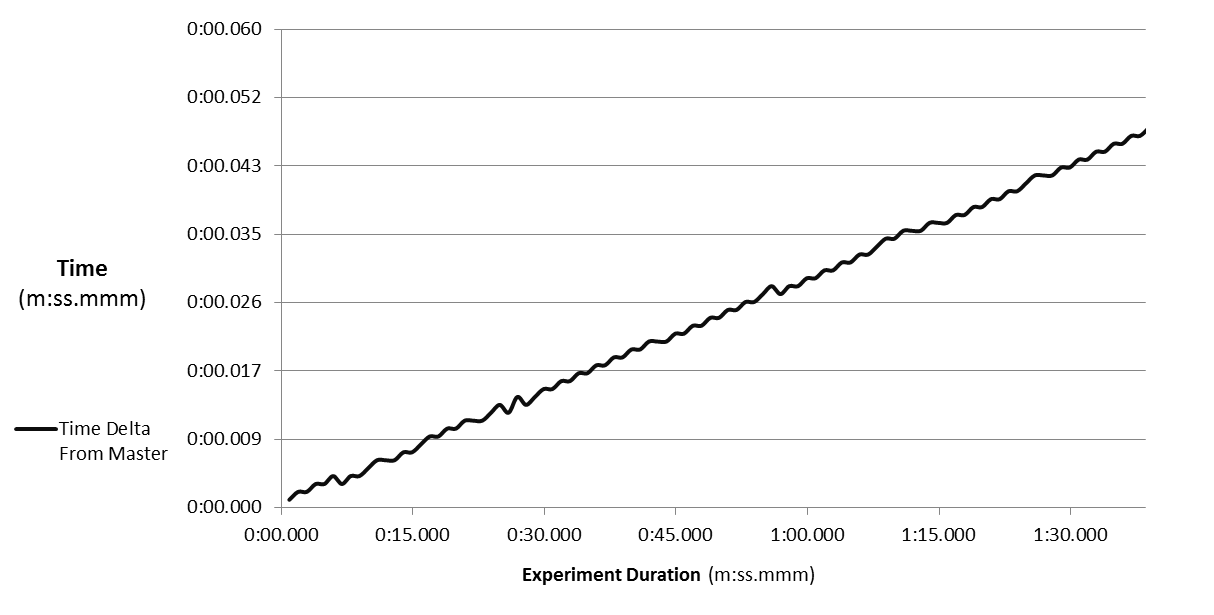}}

\caption{Hostile \texttt{DELAY_RESPONSE} messages gradually shifting the target's time away from its master. Using the default PTPd thresholds we see that the adversary can create a 26 msec discrepancy after 1 min and 90 packets sent.}

\label{Delay Spoofing}

\end{figure}

\paragraph{Mitigation}\label{time:scenario1mit}

The most basic mitigation for this attack is to add thresholds on the legal message delay ratio, as suggested in \cite{Delay_Filter_1}:

\begin{itemize}

  \item Maximum time drift per second (default: 0.5 msec/s )

  \item Upper limit on network delay measures (default: not checked)

\end{itemize}

As can be seen in Figure \ref{Delay Spoofing}, these techniques are already implemented in PTPd, resulting in extended attack efforts to create even a small time change at the target. With these countermeasures the adversary's ability to control the target's time is limited and demands a large effort, both in time and in message count.

\subsubsection{Applicative Sync Spoofing Attack Scenario}\label{time:scenrio2}

\paragraph{Attack}

The same weak adversary can mount a much more effective attack just as easily - and this attack is not dedicated against PTPd. Instead of targeting the \texttt{DELAY_RESPONSE}, the OOB Applicative adversary will send spoofed \texttt{SYNC} messages with hostile timestamps on behalf of the grandmaster, to a chosen target node. The targeted node will adopt these hostile timestamps and will recalibrate itself according to the specified attacker's time values. Note that \texttt{SYNC} is uni-directional, thus the attacker does not need to ``bend'' any traffic toward itself.

\begin{figure}[t]

\centerline{\includegraphics[width=4.5in]{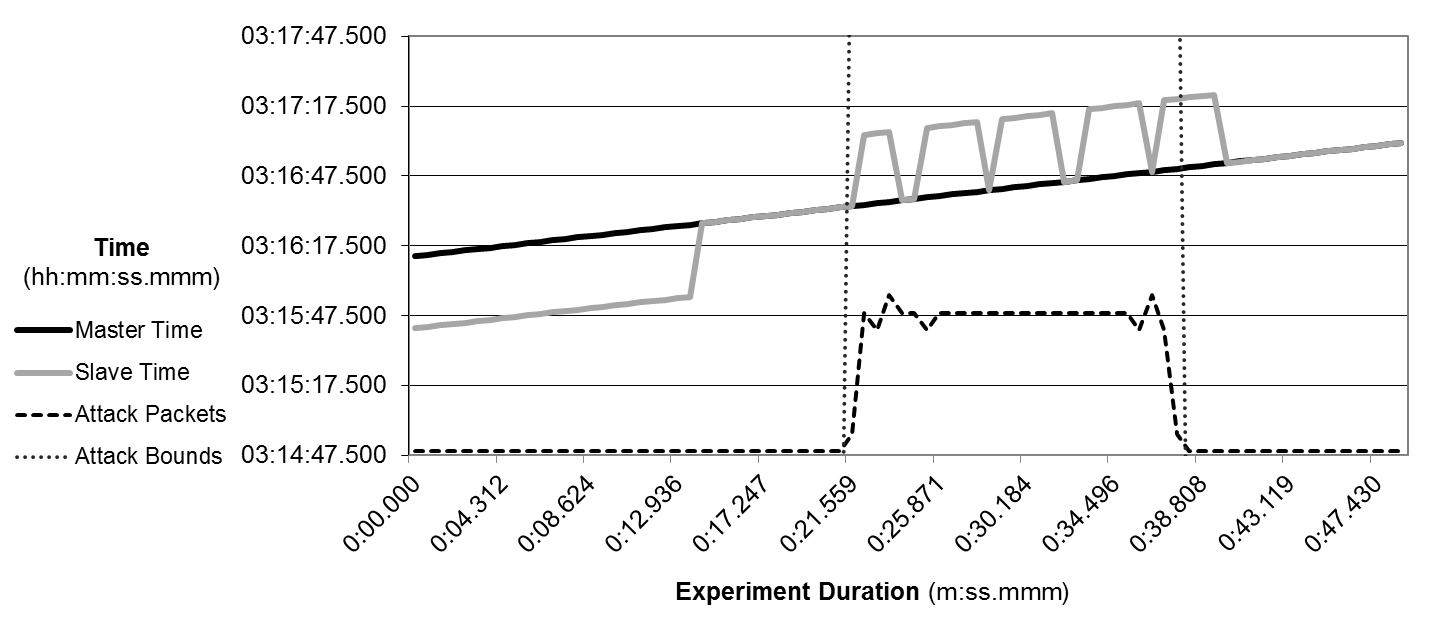}}

\caption{Applicative Sync Spoofing: Between times 0:21 and 0:37 the adversary is sending hostile \texttt{SYNC} messages, with timestamps of +30 seconds.}

\label{App Syncs}

\end{figure}

Figure \ref{App Syncs} shows what happens when the attacker sends bogus \texttt{SYNC} messages; the slave accepts the false timestamps and sets its clock to the attacker's chosen time. Once every 5 sec the legitimate grandmaster's messages reset the slave back to the true time. Note that the +30 sec value is illustrative; we successfully set the clock 20 years back.

Using thresholds on the messages, such as maximal change before reset, as done in the PTPd implementation, makes this attack slightly less effective. As can be seen in Figure \ref{App Syncs}, the PTPd's basic threshold is 1 second, and so when the true grandmaster sends his \texttt{SYNC} messages the large delta causes the slave node to reset back into the original time.

\begin{figure}[t]

\centerline{\includegraphics[width=4.5in]{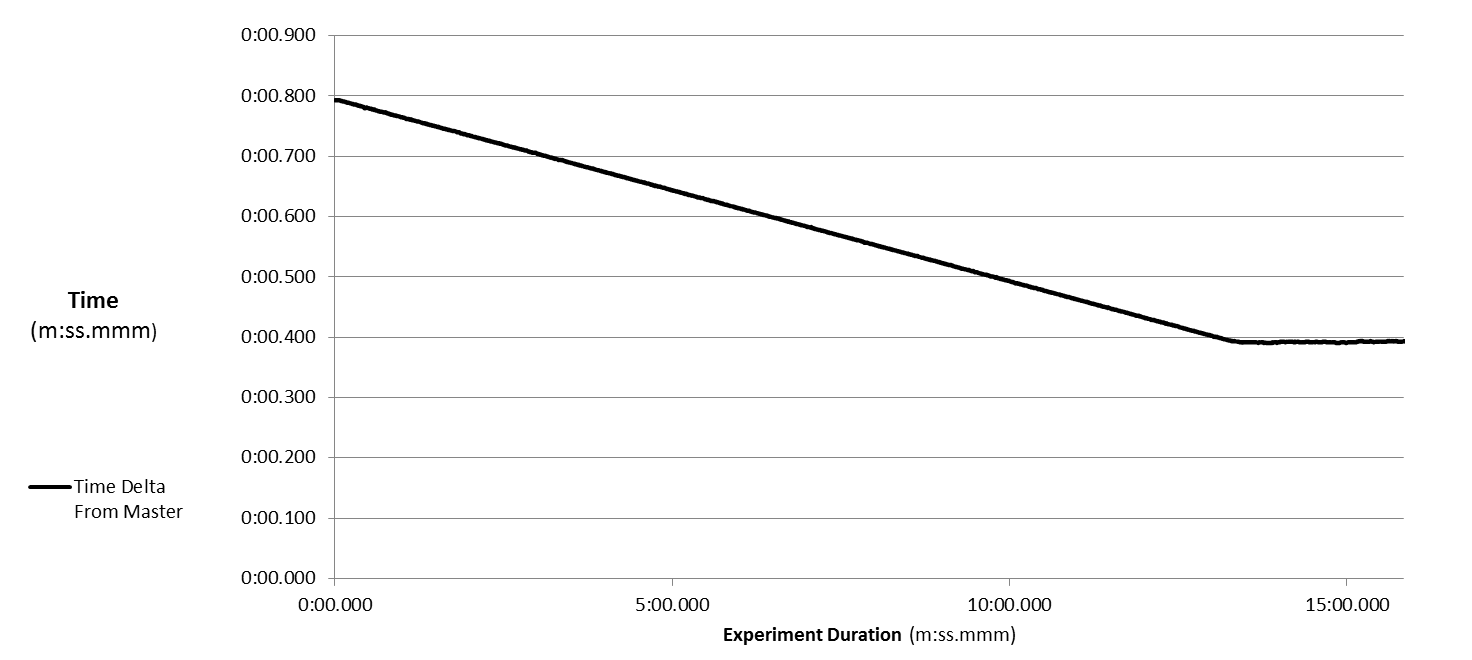}}

\caption{Duplicate Masters: OOB Applicative adversary sending hostile \texttt{SYNC} messages, with timestamps of +800 msec, converge the target to an average of +400 millis.}

\label{Dup Masters}

\end{figure}

When the attacker chooses to use smaller time differences, below the PTPd's 1 second threshold, we get a different behavior. As we can see in Figure \ref{Dup Masters}, we can observe similar results to \cite{Delay}, in which the target's time converges to the \emph{average} between the real master and the adversary's hostile time (that was higher by 800 msec).

As was shown in \cite{first_deploy}, deployment issues can easily cause similar ``attack-like'' behavior, since the PTP slave has no sense of the legitimacy of the incoming \texttt{SYNC} messages.

\paragraph{Mitigations}\label{time:scenrio2mit}

The major design flaw that enables the \emph{Sync Spoofing} attack is the fact that there is no binding between the PTP entity (the clock ID) and the underlying network ID. If such a binding were used, than the slave node would have been able to overcome the attacks using the following checks, ordered by the attack scenarios:

\begin{enumerate}

  \item Sending messages to the master's network address as derived from its clock ID

  \item Checking for a match between the PTP Clock ID and the network ID and discarding mismatching messages

\end{enumerate}

\begin{figure}[t]

\centerline{\includegraphics[width=3.5in]{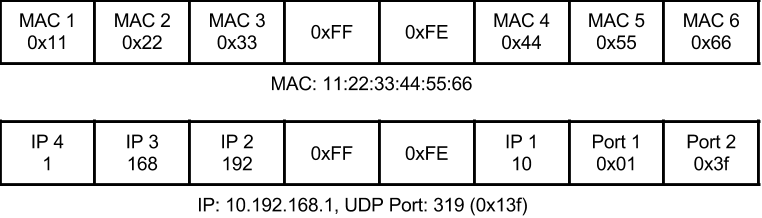}}

\caption{Construction of the clock ID from the underlying network address}

\label{clock ID}

\end{figure}

Instead of using a IEEE EUI-64 assigned clock IDs, we suggest a logistically simpler solution that maintains the original goal - network unique identities. We suggest to deterministically construct the PTP clock IDs from the underlying network IDs, as can be shown in Figure \ref{clock ID}. Note the use of the constant 0xFFFE in the middle of the clock ID, as specified in the IEEE 1588 standard, thus we maintain backward compatibility.

\begin{figure}[t]

\centerline{\includegraphics[width=4.5in]{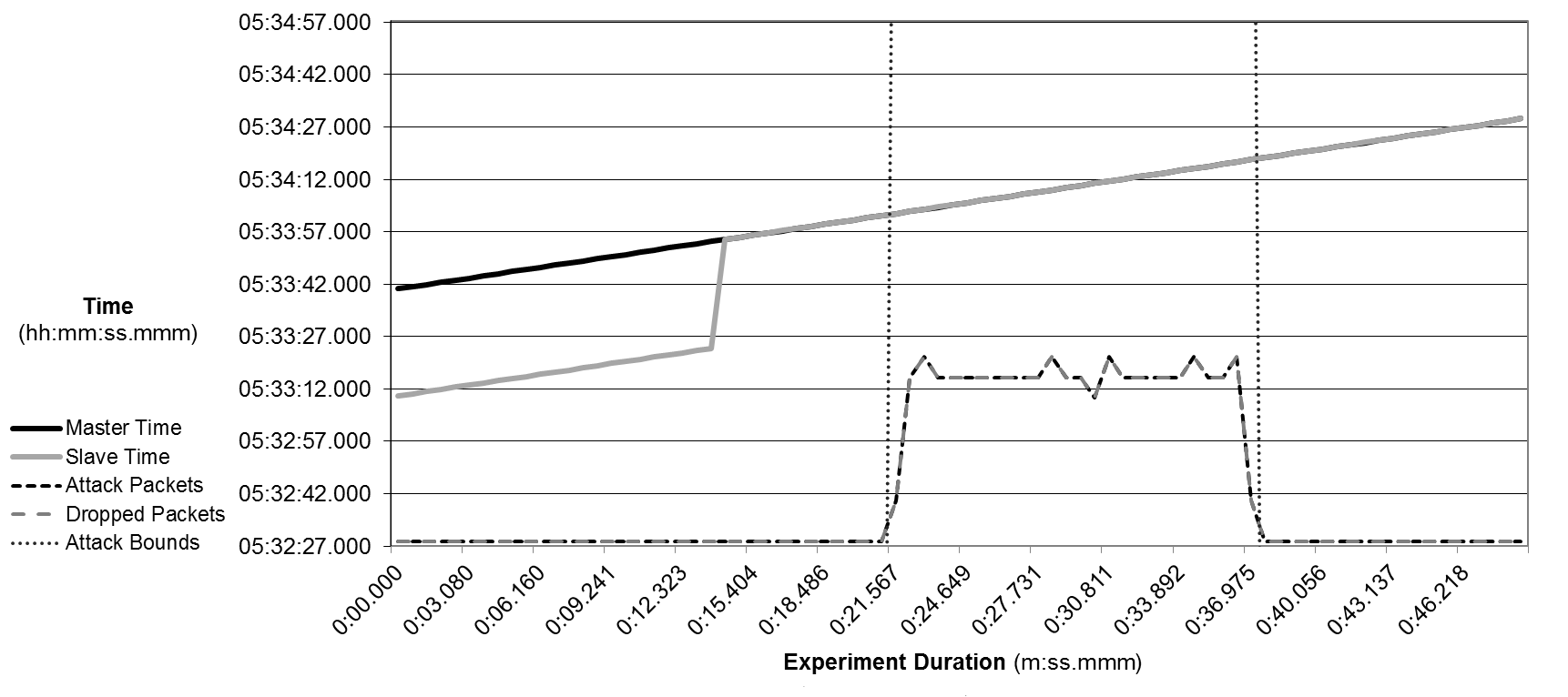}}

\caption{Applicative Sync Spoofing Mitigation: Binding between network ID and Clock ID, completely blocks the attack.}

\label{2nd Mitigation}

\end{figure}

An alternative approach makes use of the fact that only the master's binding should be publicly known. This is because the master always sends multicasts or replies to messages of the sending slave, and needs no a priori knowledge of the slaves' network address. The idea is to add the master's network address to its \texttt{ANNOUNCE} message, now resulting in his network address being publicly known to all listening PTP slave nodes. In our experimentation we chose to implement the 1st solution. Figure \ref{2nd Mitigation} shows the results when using the new defense mechanism against the same attack shown in Figure \ref{App Syncs}: The defense fully blocks the attack scenario from Section \ref{time:scenrio2}.

\subsubsection{Network Sync Spoofing Attack Scenario}\label{time:scenrio3}

\paragraph{Attack}

\begin{figure}[t]

\centerline{\includegraphics[width=4.5in]{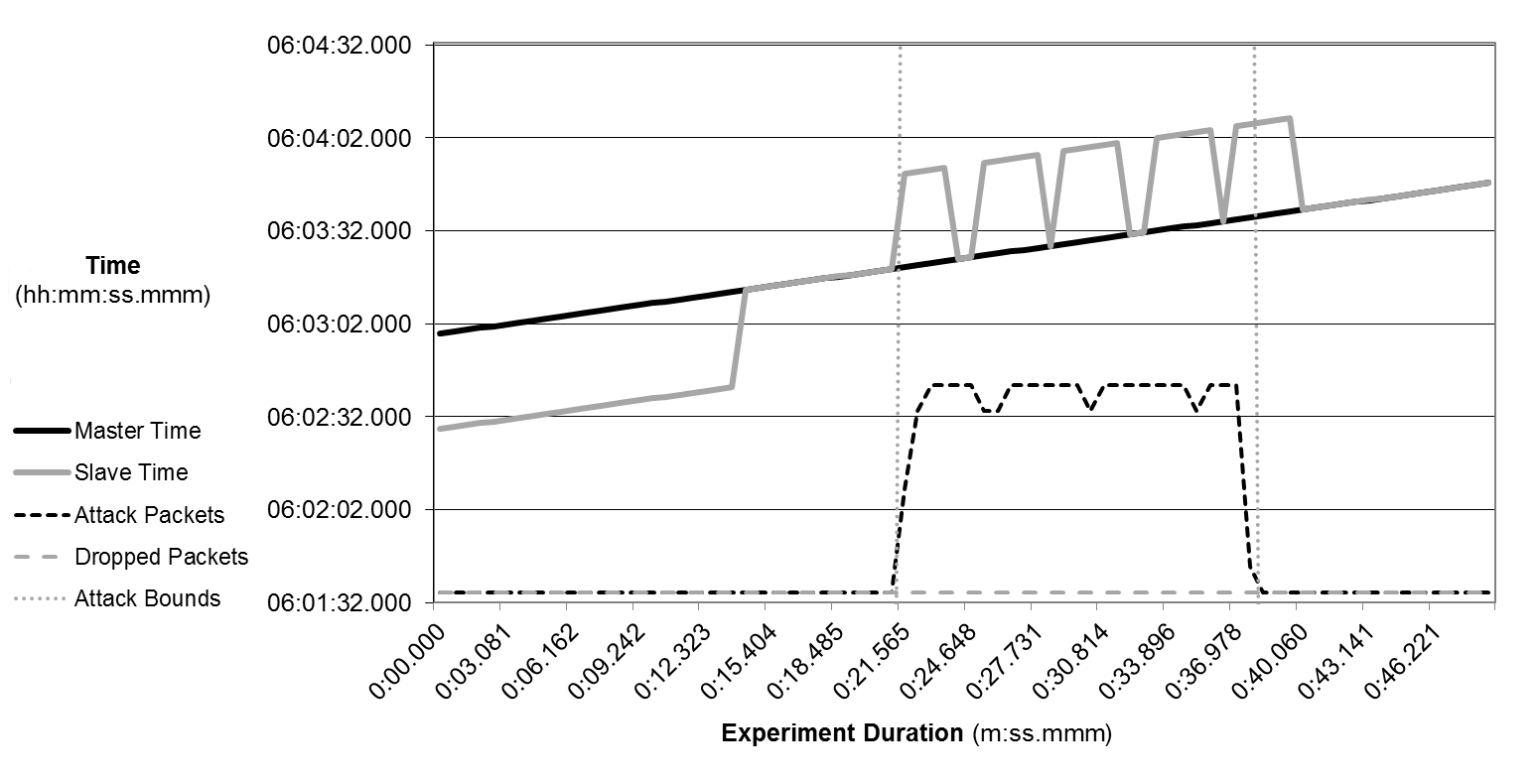}}

\caption{Network Sync Spoofing: The adversary is sending hostile \texttt{SYNC} messages, with timestamps of +30 seconds. The slave check the binding between the clock ID and the network ID - but those are spoofed by the attacker.}

\label{Net Sync Spoof}

\end{figure}

While binding the Clock ID to the network ID does offer a defense against the weakest attacker, a slightly stronger attacker (OOB Network Adversary) can still deploy the \emph{Sync Spoofing} Attack of Section \ref{time:scenrio2}: all it needs to do is spoof the underlying network addresses. Figure \ref{Net Sync Spoof} shows a Sync Spoofing attack using IP spoofing towards the master.

\paragraph{Mitigations}\label{time:scenrio3mit}

The adversary is making use of the fact that the messages from the master to the slave are treated independently and without any session semantics. Although the PTP header defines the \texttt{sequenceId} field, it is only used for distinguishing the messages from one another. More explicitly, the standard does not define the following issues regarding a \texttt{sequenceId} field:

\begin{enumerate}

  \item What is the initial value of the sequence ID of a given message?

  \item Should the receiver check the a connection between two consecutive messages' sequence IDs?

\end{enumerate}

The resulting effect is that the receiving node can not check the validity of the \texttt{sequenceId} field of the received message, as the validity is not properly defined.

We suggest to extend the standard and make use of the \texttt{sequenceId} field for adding session semantics between the master node and the slave node:

\begin{enumerate}

  \item Master originating messages, \texttt{SYNC} and \texttt{ANNOUNCE}, will start with a pseudo-random sequence ID counter (to prevent session hijacking), each message type will have a separate counter

  \item A message's sequence ID will be incremented by 1 for each sent message

  \item The slave node will validate a received message's sequence ID, and check that it matches a predefined range of available ids, compensating for a possible network errors. The range of counter values acceptable to the slave is called the \emph{Window}.

\end{enumerate}

In addition, the suggested extension can be used to strengthen the delay/response mechanism of the protocol:

\begin{enumerate}

  \setcounter{enumi}{3}

  \item Each \texttt{REQUEST_<X>} message from the slave to the master will have a pseudo-random sequence ID, acting as a challenge to the master

\end{enumerate}

It is worth mentioning that while these session semantics strengthen the connection between the master node and the slave node, the \texttt{SYNC} messages are still sent in multicast and so are, by definition, publicly accessible to the OOB attacker. This can be solved by maintaining a separate \texttt{sequenceId} for each slave node, as specified by the standard, and sending the \texttt{SYNC} messages in unicast. The unicast registration and transmission mechanism is already defined in the standard, and was originally designed for Annex K's deployment. Due to its heavy costs, the transition to unicast transmission should be applied only in network scenarios in which there is no separation between broadcast messages and multicast messages, such as PTP over Ethernet. In our implementation we used PTP over UDP and did not use the unicast transmission.

\begin{figure}[t]

\centerline{\includegraphics[width=5in]{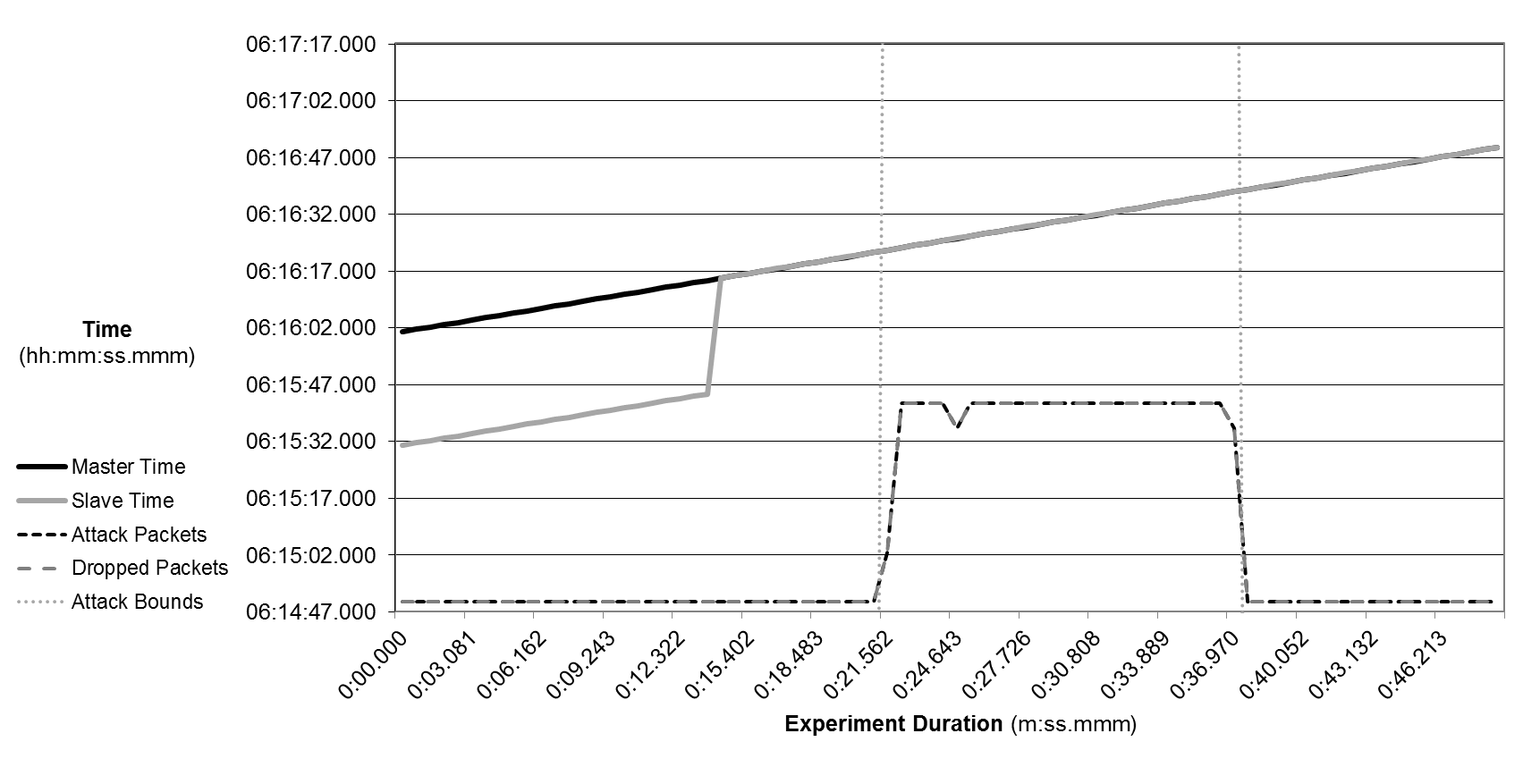}}

\caption{Network Sync Spoofing Mitigation: Using the unknown 16-bit \texttt{sequenceId} values as a session binding completely blocks the attack: we can see that all the attack packets are dropped.}

\label{Def Net Syncs}

\end{figure}

Applying these defense mechanisms forces the OOB attacker to send his spoofed messages with a correctly guessed 16-bit secret, thus significantly lowering his success rate, as can be seen in Figure \ref{Def Net Syncs}.

\subsubsection{Blind Window Snatching Attack Scenario}\label{time:scenrio4}

\paragraph{Attack}

16-bit sequence number only offers limited protection. The attacker can use two ideas to give him an advantage: (a) he can transmit much faster than the real grandmaster, and (b) he can use the defender's Window against him, by incrementing the counter by a whole window with each packet. Using these ideas, the OOB Network adversary can deploy a \emph{blind window snatching} attack, after which he will be able to deploy the attack from Section \ref{time:scenrio3}. The \emph{blind window snatching} attack makes use of the PTP window behavior: after a valid message is received, the window automatically advances to start from the message's \texttt{sequenceId} + 1. This is due to the fact that the protocol messages are time oriented and so, unlike the TCP window, if for example the slave receives a message with a \texttt{sequenceId} of 5 instead of 4, there is no need to wait for the ``old'' message with the lower \texttt{sequenceId} 4, as it is no longer interesting. Listing 1 shows code implementing the attack.

\renewcommand{\ttdefault}{pcr}

\begin{lstlisting}[float,floatplacement=t,
					frame=single,
					language=Python,
					basicstyle=\ttfamily,
					keywordstyle=\ttfamily\bfseries,
					commentstyle=\ttfamily\itshape\color{gray!90},
					caption={Python code for the \emph{blind window snatching} attack}]
##
# M - A pre-formatted SYNC message
# W - The size of the window
# R - The full ID range [0, 65535]
##

def blind_window_snatch(M, R, W):
    # Initialized to the top of the first window
    seqId = W - 1
    # Split the range R to window (W) sized subranges
    for i in range(R / W - 1):
        # send the upper most sequenceId in the given subrange
        sendmsg(M, i * W + seqId)
    # The window has wrapped-around and 0 is now that start of it
    seqId = 0
\end{lstlisting}

\renewcommand{\ttdefault}{cmtt}

\begin{figure}[t]

\centerline{\includegraphics[width=4.5in]{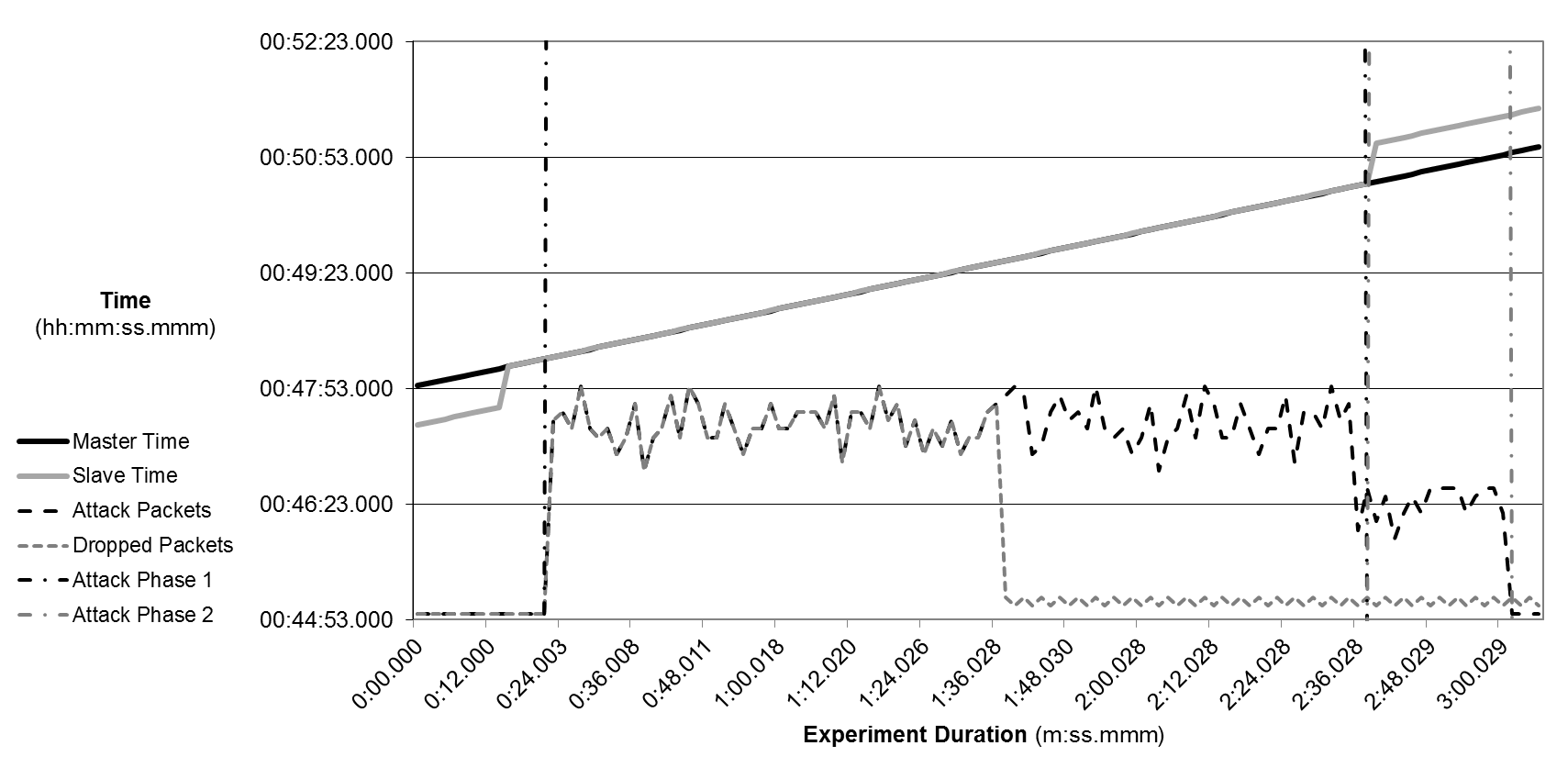}}

\caption{Phase I: performing a blind window snatching attack on window of size 50. Sending packets at a rate of 10 pps, the attack goes through 65536/50 ($\approx$ 1300) packets in slightly more than 2 minutes. Phase II: Sync Spoofing.}

\label{Win shift fig}

\end{figure}

What this attack does is snatch the window from the true master. At some point the attacker's packet will hit the legitimate window, and increment it - so all subsequent legitimate packets fall outside the new window and are discarded. As can be seen in Figure \ref{Win shift fig}, once the adversary ``catches'' the target's session window (at time 1:36), all the master's messages fall outside the window, resulting in logged message drops.

This means that the total message complexity of the blind window snatching attack on a target with \texttt{sequenceId} from range $R$, a window size of $w$, and an original attack with message complexity of $K$ is: 

\begin{equation} \label{WindowComplex}
\#messages=\frac{|R|}{w} + K
\end{equation}

This is a major improvement compared to the naive attack in which the original attack ($K$ messages) is preformed for every possible window:

\begin{equation} \label{NaiveComplex}
\#messages=\frac{|R|}{w} \cdot K
\end{equation}

An added bonus of this attack is that the adversary now completely ``owns'' the session window, winning him a quiet environment without interference caused by the master's messages, as they all fall outside the window.

There is a small chance that the window will wrap-around during the attack, resulting in the attack failing to catch the window. This can easily be handled by performing the window snatching procedure twice: the high send rate of the adversary prevents the wrap-around from happening during the second window snatching. The updated message complexity will now be:

\begin{equation} \label{FinalWindowComplex}
\#messages=2 \cdot \frac{|R|}{w} + K
\end{equation}

In case that the attack was successful on the first attempt, the second repetition only shifts the already controlled window.

\paragraph{Mitigation}\label{time:scenrio4:mit}

The blind window snatching attack depends on the rather small 16-bit range $R$ caused by the 2 byte size of \texttt{sequenceId}. We suggest to overcome this limitation by enlarging the \texttt{sequenceId} field using 2 of the reserved bytes (out of the 5.5 currently reserved bytes in the message's header). By using a 4 byte sized \texttt{sequenceId} field, even if the \emph{blind window snatching} attack will send 1000 messages per second and the window will be of size 16, it will still take the adversary more than 3 days per snatch.

\begin{table}[t]

\centerline{\includegraphics[width=3in]{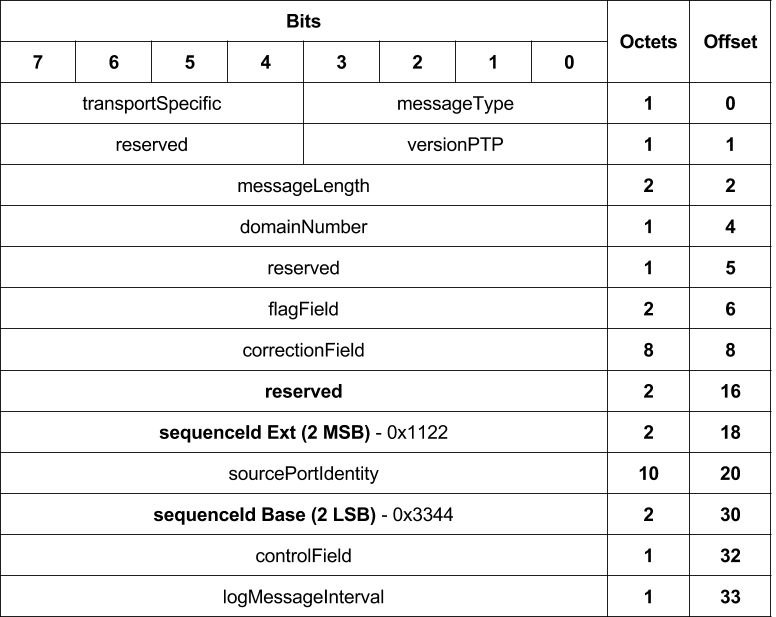}}
\caption{Using reserve bytes for the extended sequence ID with value: 0x11223344}
\label{PTP Header update}

\end{table}

In Table \ref{PTP Header update} we show a possible updated header's structure. By storing the 2 MSB in the reserved bytes, and using a reserved flag from the \texttt{flagField}, to mark the 4 byte support, one can gradually deploy the upgraded version while maintaining backward compatibility to the current 2 byte field. This will result in a node's keeping an additional boolean variable in his \texttt{parentDS}. This variable will store the master's version compatibility, and will be used for checking consistency between master messages, as there is no reason that the master's compatibility will change during the session. Note that the reserved bit from the \texttt{flagField} can be replaced using the existing \texttt{SECURITY} bit, ment originally for Annex K.

\subsubsection{Inband Sync Spoofing Scenario}\label{time:scenrio5}

\paragraph{Attack}

The In-Band adversary can also perform the \emph{Sync Spoofing} attack described in Section \ref{time:scenrio3}. The difference is that the adversary has a MITM position that enables him to sniff the session ``secrets'', therefore the defenses of Section \ref{time:scenrio3} and \ref{time:scenrio4} are ineffective.

\paragraph{Mitigation}\label{time:scenrio5mit}

Because the adversary is In-Band, the only way to withhold secrets from him is by using a cryptographic solution. Annex K suggests using a symmetric key approach, which would in fact defend against an In-Band outsider attack. However, we suggest to utilize a Public-Key cryptographic solution, that can defend against much stronger (Insider) attacks as well.

\subsubsection{PTP Insider Threat}\label{time:scenrio6}

\paragraph{Attack}

Assuming Annex K is activated with its symmetric-key defenses, a hostile PTP node can pretend to be the master node and initiate the 3-way handshake against the target slave. By using the legitimate symmetric key, the handshake will complete successfully resulting in the slave being fooled to think that the adversary is the legitimate master node. From now on, the adversary will use the derived symmetric key to send hostile \texttt{SYNC} messages to his target.

\paragraph{Mitigation}\label{time:scenrio6mit}

This attack scenario addresses a major flaw in the standard's Annex K: there is no way to differentiate between the slave nodes and the master nodes. This leads to an inherent vulnerability when addressing an insider threat. Variants of the described attack will work against any other non-role based defense scheme deployed in the network, including the use of standard solutions such as IPsec or MACsec as was suggested in \cite{ISPCS_2012}.

We suggest a different approach, that will focus on a role-based solution that will prevent a slave node from masquerading as the master node. The proposed solution is to replace Annex K in favor of Public-Key based cryptography. By equipping each of the master-capable PTP nodes with a unique private/public key pair, we enable the elected master node to use the \texttt{ANNOUNCE} message to publicly declare his public key. From now on, the master node will sign its \texttt{SYNC} messages, and the slave nodes will be able to use the master's public key to verify the legitimacy of the received timestamps. We provide more details, and a performance evaluation of the suggested solution in Section \ref{Solution}.


\subsection{BMC Attacks}\label{attack:bmc}

The IEEE 1588 standard specifies the BMC algorithm as a leader election that elects the network's grandmaster clock in a distributed fashion. The chosen grandmaster clock is responsible for the synchronization of the PTP slave nodes. This powerful vantage point makes the BMC algorithm a desired target for an adversary.

In this section we  focus on the attack scenarios that target the BMC algorithm itself.

\subsubsection{Outsider Rogue Master Attack Scenario}\label{bmc:scenrio1}

\paragraph{Attack}

The weakest possible adversary, an OOB applicative adversary, can propose himself as a grandmaster candidate by sending fake \texttt{ANNOUNCE} messages declaring him to be the best clock in the network. This can easily be done by faking to be a truly magnificent clock: Using the minimal value for the \texttt{Priority1} and \texttt{Priority2} fields; declaring that the clock's source as \texttt{ATOMIC_CLOCK}; etc.

\begin{figure}[t]

\centerline{\includegraphics[width=4.5in]{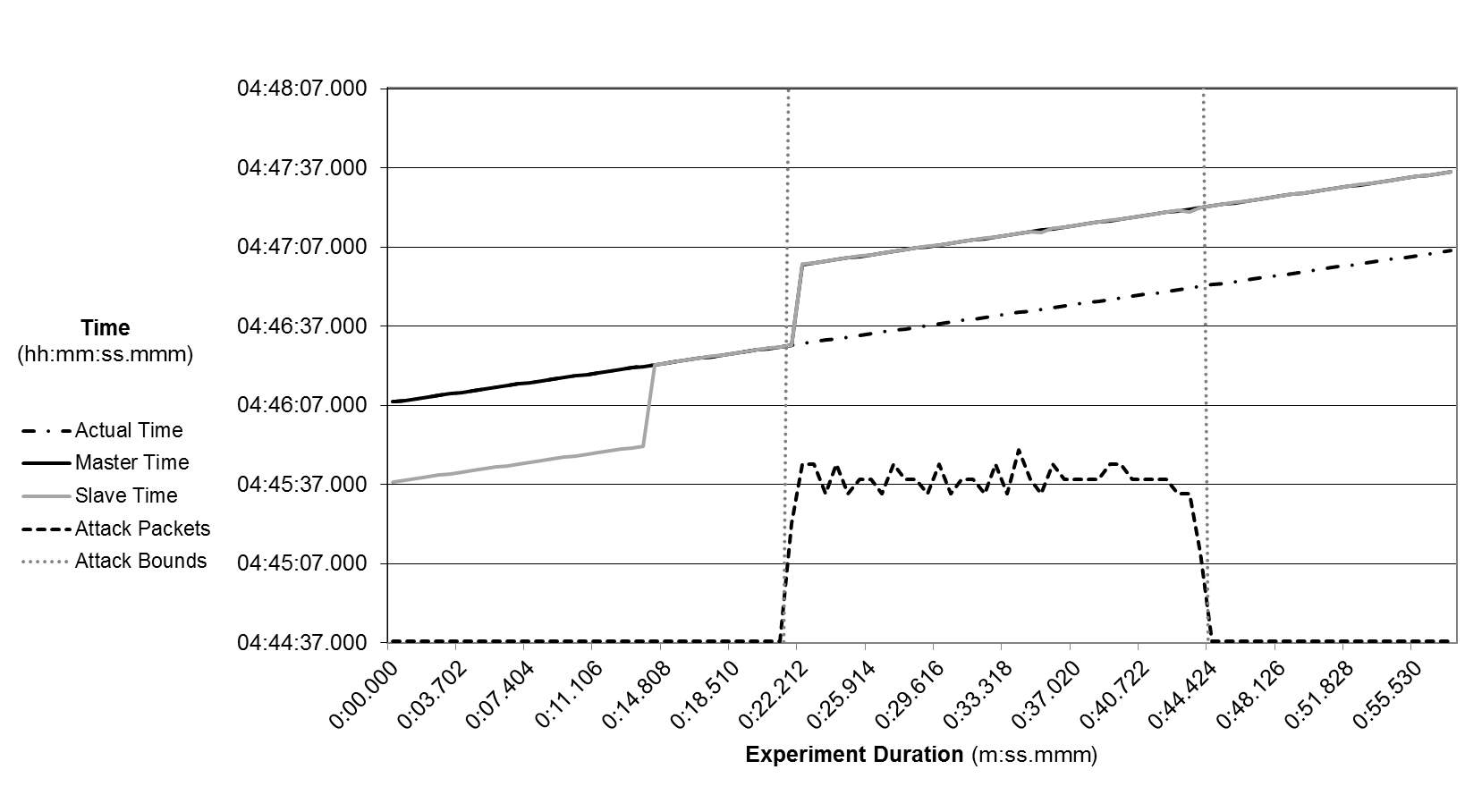}}

\caption{Rogue Master: An outsider adversary sends \texttt{ANNOUNCE} messages to nominate himself to be the grandmaster.}

\label{Rogue Master}

\end{figure}

We implemented this attack and tested it. As can be seen in Figure \ref{Rogue Master}, even after the adversary stops his attack (at time 0:44), the network will continue on according to his hostile time: this is because the original grandmaster will now dictate the adversary's hostile timestamp to everyone on the network.

\paragraph{Mitigation}\label{bmc:scenrio1mit}

In a similar way to the problem shown in Figure \ref{App Syncs}, the problem is that there is no separation between the \emph{Outsider} identities in the physical network and the \emph{Insider} identities in the PTP network. This can be addressed by the use of Annex K, or by our own Public-Key-based solution.

\subsubsection{Insider Rogue Master Attack Scenario}\label{bmc:scenrio2}

\paragraph{Attack}

If a hostile PTP node proposes himself as a grandmaster candidate by sending fake \texttt{ANNOUNCE} messages declaring him to be the best clock in the network - Annex K, and any Symmetric-Key solution, will fall short of defending against an insider threat. This is due to the fact that the insiders have the secret keys and can appear as a convincing grandsmaster.

\paragraph{Mitigation}\label{bmc:scenrio2mit}

The solution here is the same as for the PTP insider threat (\ref{time:scenrio6}): use public key cryptography.

We implemented such a tailor made public-key cryptographic solution and observed that it works without problems, unlike the unsuccessful use of ordinary VPN based solutions (\cite{nokia}).


\subsection{Management Layer Attacks}\label{attack:management}

The IEEE 1588 standard \cite{standard} defines a PTP specific management format. The format enables the management node to query PTP nodes' dataset fields, using \texttt{GET_<X>} messages, and update them, using \texttt{SET_<X>} messages. For example, using the \texttt{SET_CLOCK_ACCURACY} message, the management node can update a specific node's accuracy field, an action that can change the result of the next BMC algorithm. Although the protocol defines extensive management capabilities, it does not specify any authentication mechanism to be used when identifying the management node. Further, note that management messages are uni-directional and do not require a response.

\subsubsection{Applicative Proxy Grandmaster Attack Scenario}\label{manag:scenrio1}

\paragraph{Attack}

Even the weakest adversary, the OOB Applicative adversary, can send fake management messages to upgrade a specified PTP node's dataset. As a result of its dataset improvement, the targeted node will win the BMC leader election and will be declared the network's grandmaster. After being elected, the adversary will send \texttt{SET_TIME} messages to directly control the target's time, thus taking full control of the time and structure of the whole PTP network.

\paragraph{Mitigation}\label{manag:socioeconomic}

The above attack takes advantage of the fact that the 2008 standard version holds no authenticated mechanisms between the management node and the ``ordinary'' nodes. The straightforward defense mechanism that can be deployed is a white-list of IPs, or other network address matching the network's underlayers, from which management messages should be sent.

\subsubsection{Network Proxy Grandmaster Attack Scenario}\label{manag:scenrio2}

\paragraph{Attack}

A whitelist or blacklist mitigation is still easily defeated; The OOB Network adversary, or any other stronger adversary, will deploy the previous attack while also spoofing the network layers of the messages so to match with the management's white-list of addresses.

\paragraph{Mitigation}\label{manag:scenrio2mit}

There are 3 possible solutions to assure the management message's authenticity:

\begin{enumerate}

  \item Adding a session binding between the management node and the ordinary nodes

  \item Adding a cryptographic authentication to the management messages

  \item Deprecating the management protocol, in favor of more standard and secure management and control protocol

\end{enumerate}

Although the first 2 solutions can be deployed, each with its pros and cons, we agree with \cite{PTP_Deploy} and recommend to pick the 3rd option. The management messages can easily be replaced by a standard SNMP implementation that natively supports authentication without needed changes \cite{SNMPv3}. SNMP maintains the Set and Get basic operations of the original management protocol, and can be implemented using PTP specific MIB structures. The main advantage of this proposed countermeasure is explained in \cite{PTP_Deploy} from a deployment's point of view.


\section{Proposed Security Extension}\label{Solution}

Up to this point we have shown the continuous arms race between the adversary and out proposed defenses. In this section we describe the full details of our cryptographic defense mechanism, its analysis compared to the original Annex K and to state of the arts security proposals, and its remaining gaps.

\subsection{Prerequisites and Scope}\label{overview:pre}

Our defense scheme assumes the existence of several prerequisites, which are:

\begin{enumerate}

  \item The existence of a management entity

  \item A predefined public verification key, distributed to all PTP nodes

  \item Master certificates signed by the management entity

\end{enumerate}

The first assumption can easily be solved in a real deployment of the protocol since there will be an administrator to set up the system. Nevertheless, the assumption is only needed for the sake of the rest assumptions, thus making the management task relatively easy.

The second assumption is the strongest assumption that we demand, and still it can be solved using ordinary management tools in the computer network. The PTP's deployment is meant to accurately distribute the time between an already defined, and probably managed, computer network. Therefore, we only reduce the problem of Public-Key distribution for the PTP protocol, to the already existing problem of Public-Key distribution for the entire computer network.

The third assumption can be solved in a straightforward manner by the management entity that can choose between 2 rather simple solutions:

\begin{enumerate}

  \item Presupplying the certificates when configuring the PTP middleboxes

  \item Acting as a distribution server that distributes existing certificates on demand

\end{enumerate}

In this paper we focus solely on the PTP header format and the following PTP messages: \texttt{SYNC}, \texttt{FOLLOW_UP}, \texttt{DELAY_REQUEST}, \texttt{DELAY_RESPONSE} and \texttt{ANNOUNCE}. The solution can easily be extended to the rest of the messages.

In addition, for simplicity's sake, the explanation below  makes an assumption that the clock IDs are generated in accordance with the underlying network addresses, as specified in Section \ref{time:scenrio2}. Otherwise, the \texttt{ANNOUNCE} message should have included an additional field declaring the node's network address, and the slave would need to store in his master's dataset this additional field.

\subsection{Cryptographic Design Choices}\label{overview:crypto}

As was suggested in \cite{Threat_Analysis}, we have implemented and demonstrated our cryptographic solutions. In our implementation we chose an Edwards-Curve (Ed) based public key scheme, and more specifically, the EdDSA signing scheme \cite{Ed}. Our choice is based on 4 main arguments:

\begin{enumerate}

  \item Only the authenticity is important, and so we need only a signing/verification scheme

  \item Ed schemes make use of relatively short keys

  \item Ed schemes make use of relatively short signatures

  \item Ed schemes are designed to be faster than other digital signatures schemes

\end{enumerate}

While the first argument does not specifically point to Ed based solutions, the rest of the arguments had an important role in our choice.

When the size of the \texttt{ANNOUNCE} message (the largest PTP message) is merely 64 bytes, the signature that is being used should try to match it in size. In our implementation over the open-sourced PTPd, we integrated an Ed25519 scheme \cite{Ed25519}, based on the \emph{WolfCrypt} library \cite{WolfCrypt}. Using this scheme the size of a signature is 64 bytes, and the size of the public key is 32 bytes, thus maintaining the rather small size of the original messages.

\subsection{PTP Header Format}\label{sol:header}

Since easing the deployment process was one of our design goals, we made every effort to keep the protocol's original header format. The only extensions we suggest to the header format were already discussed in Section \ref{time:scenrio4:mit}:

\begin{enumerate}

  \item Using 2 reserved bytes for the enlarged 4 byte \texttt{sequenceId}

  \item Using 1 reserved bit, from the bit flag, marking the use of the defense scheme

\end{enumerate}

While the enlarged \texttt{sequenceId} enables session-like semantics between the master and the slave, it also acts as a cryptographic aid: an effective anti-replay counter. By making the slave check the window validity of this field, and by incrementing it slowly through the large 32 bit range, the adversary's ability to replay a recorded signed message is practically non-existent. In case the master sends 20 \texttt{SYNC} messages per second, the identifier's wrap-around will occur after approximately 7 years.

The second use of the enlarged \texttt{sequenceId} field, as was briefly described in Section \ref{time:scenrio3mit}, is as a challenge-response field for the delay/response messages. Meaning that before the slave sends a \texttt{DELAY_REQUEST} message, it randomly picks a \texttt{sequenceId}, therefore generating a challenge to the waiting adversary. While the master node simply needs to reply with the same \texttt{sequenceId}, an OOB adversary will need to guess the challenge. In addition, in case these messages were to be signed too, even the In-Band adversary would need to store a large amount of recorded signed \texttt{DELAY_RESPONSE} messages for him to even try to effect the delay measurements (the adversary can utilize the Birthday Paradox to reduce his storage costs).

\subsection{Announce message Format}\label{sol:announce}

The original design of the IEEE 1588 standard, allows flexibility in the structure of the \texttt{ANNOUNCE} message. This allows us to use the \texttt{ANNOUNCE} as a message that wraps together all of the node's specifications, including the distribution of cryptographic certificates throughout the PTP network.

\begin{table}[t]

\centerline{\includegraphics[width=3in]{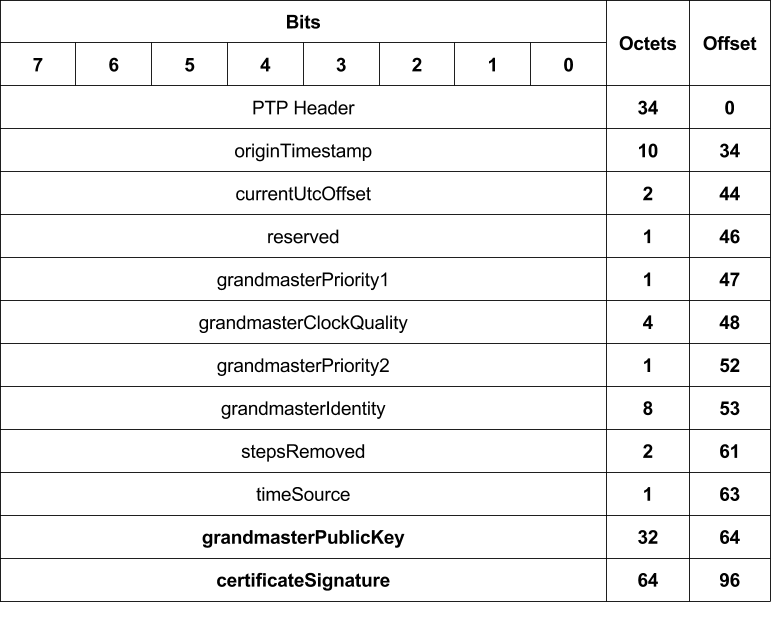}}

\caption{Adding the certificate to the announce message: public key + signature}

\label{announce message}

\end{table}

The required change to the message's structure is only the addition of two extension fields, as can be seen in Table \ref{announce message}.

The first field is the master's public key, to be stored and used by slave nodes. In our implementation this field had the size of 32 bytes.

The second field is the management's signature over all of the \texttt{ANNOUNCE} fields, including the public key. In our implementation this field had the size of 64 bytes. This results in an \texttt{ANNOUNCE} message of size 160 bytes, instead of the original 64 bytes.

\textbf{Technical Note \#1:} The signature covers only the \texttt{ANNOUNCE} message's body, without the PTP header. In addition, some of the message's fields are not needed for the BMC algorithm and so they are filled with 0 when the certificate is created. These fields are: \texttt{originTimestamp}, \texttt{reserved} and \texttt{stepsRemoved}.

\textbf{Technical Note \#2:} During the BMC algorithm, after a slave validates a master candidate's certificate once, it needs only to compare its fields and signature with every newly received \texttt{ANNOUNCE} message from the same node. Only in case the certificate parameters change is there a need to actually perform a cryptographic verification check on it again. This means that the expectation of the overhead cost in signing and verifying these messages is practically zero: The management node needs to sign the certificate once, the master node only needs to copy it from its configuration, and the slave node needs only check it once per BMC-elected master.

\subsection{Sync message Format}\label{sol:sync}

The \texttt{SYNC} message stayed unchanged in our scheme, due to the fact that we mandate the \texttt{FOLLOW_UP} message to allow hardware-based timestamping. The \texttt{SYNC} message only acts as a time bookmark for the following message that holds the actual timestamp and signature. This protocol enabled feature, makes the addition of software based security layer possible. Meaning that the CPU needs to query the network adapter for the sent \texttt{SYNC} exact timestamp, and from there on it is free to do as it likes when building the matching \texttt{FOLLOW_UP} message.

\subsection{Follow Up message Format}\label{sol:fu}

\begin{table}[t]

\centerline{\includegraphics[width=3in]{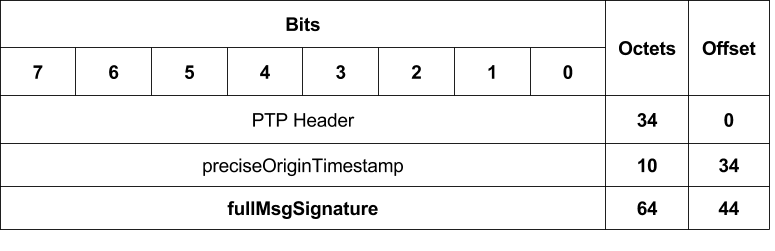}}

\caption{Adding the signature to the follow up message}

\label{follow up message}

\end{table}

The \texttt{FOLLOW_UP} message was only extended with the signature at its end. As we can see in Table \ref{follow up message}, the \texttt{FOLLOW_UP} message size grew in our implementation from 44 bytes to 108 bytes. Note that the signature is calculated over the entire PTP message: header and payload.

\subsection{Delay request and response}\label{sol:delay}

In our implementation we chose to keep the \texttt{DELAY_REQUEST} and \texttt{DELAY_RESPONSE} message untouched, i.e., not to sign the \texttt{DELAY_RESPONSE} message. This was done because unlike the \texttt{ANNOUNCE} and \texttt{SYNC} that are sent in multicast, these message are sent in unicast, resulting in an overhead on the master proportional to the number of slaves in the network. Nevertheless, in the basic PTPd rate of 1 request message per second, the cryptographic overhead is pretty minor even with 100 PTP slave nodes. However, adding a signature to the \texttt{DELAY_RESPONSE} message will cause the master to preform a cryptographic calculation after each \texttt{DELAY_REQUEST} message he receives. This can easily enable an adversary to prefrom a Denial-Of-Service (DoS) attack on the master.

An additional argument against the signing of these messages, is the rather low benefit one can gain from it. The \texttt{sequenceId} changes practically closed all attacks from OOB adversaries, while even the gap remaining can easily be handled by deploying several sanity thresholds on the received timestamps, as suggested in \cite{Delay_Filter_1}, \cite{Delay_Filter_2} and \cite{game_theoretic}. For example, although the message includes a 10 byte timestamp, for any sane deployment the timed delay probably won't exceed 1 hour.

\subsection{Performance}\label{overview:perf}

During our experimentation we discovered that the additional performance incurred from adding the cryptographic layer was very small. Our experiments were done using the \emph{DeterLab} project environment, on which we used computers with specs of: 4 X Intel(R) Xeon (R) CPU X3210 @ 2.13Ghz, 4096 Bytes cache. According to our measurements, the average cost of signing the \texttt{FOLLOW_UP} 44 bytes message was only 0.41 msec. The cost of verifying that message by the slave was only 0.17 msec. This means that even at a high rate of 128 messages per second, suitable for the telecom profile, the cryptographic work done by the CPU is negligible compared to the rest of the CPU needed calculations. These positive results show that no special hardware is needed for the cryptographic layer, as was suggested in \cite{ISPCS_2012} and implemented in \cite{FPGA}, due to the lightweight bandwidth of the PTP protocol.

\subsection{Comparison to Annex K and State of the Art}\label{sol:appendix}

There is almost no connection between our security proposal to the original Annex K. The appendix uses a symmetric cryptographic solution to establish unicast trust relationships between two PTP nodes. This means that the appendix can not handle an \emph{Insider} threat, and is specifically vulnerable to the attacks described in \ref{time:scenrio6} and \ref{bmc:scenrio2}. In addition, the appendix requires a pretty strong starting point of a preshared symmetric key, and has a significant communication overhead due to the 3-way handshake.

Our solution overcomes these problems due to its asymmetric nature: we suggest only a one way trust relation: slaves authenticate the master. Together with authenticated BMC algorithm, this enables a lightweight multicast solution, that is specifically tailored to the PTP protocol.

Recalling the 4-Pronged approach of \cite{Prongs} we see that our security extension has all of the needed qualities, but it achieves them in a different way than what was proposed before:

\begin{itemize}

  \item Message integrity - asymmetric signature (EdDSA) originating in the master

  \item Anti-Replay - using the protocol's existing \texttt{sequenceId} field, after enlargement to 4 bytes

  \item Security association - binding the network identity to the clock ID, and using the clock ID as the security ID

  \item Key Distribution - predefined management public key, and predefined master-candidates certificates (including the public keys)

\end{itemize}

In addition, we achieved additional security goals, goals that were yet to be addressed:

\begin{itemize}

  \item Node Authorization - signed \texttt{ANNOUNCE} certificates for the BMC algorithm

  \item Challenge Response - using the protocol's existing \texttt{sequenceId} field, as a challenge response for the delay/resp messages

\end{itemize}

And last, we suggest to deploy a standard and secure management protocol (e.g. SNMP), as a replacement to the PTP specific management. This protocol will be the \emph{Prong D} layer of the IEEE 1588 standard.

\subsection{Remaining Gaps}\label{sol:gaps}

After ensuring the correctness of the BMC algorithm, the authenticity of the management functionality and the validity of the protocol's timestamps, there are still some remaining gaps. These remaining gaps are against an In-Band adversary, or against the 2 all powerful insider adversaries we mentioned at the beginning: The In-Band adversary can simply add a wanted delay to the original message's network path. This attack will result in the same effect as encoding the same delay into the message's timestamp field, while no cryptographic solution will be able to protect against the described attack scenario. Suggested mitigations against such attack involve the deployment of multiple network paths, as in \cite{duplicate_PTP}, \cite{multiple_paths} and \cite{byzantine}, or using the threshold defense mechanism.


\section{Conclusions and Future Work}\label{Conclusions}

In our work we have shown a detailed threat analysis of the Precise Time Protocol, including several new attacks and mitigations. In the analysis we demonstrated the main security vulnerabilities of the protocol: naive BMC leader election, master timestamps without verifications, and lack of management authentication. Although the protocol defines the security Annex K, this appendix falls short of solving all of the protocol's security issues. The appendix makes use of a pre-shared symmetric key, and defines a three-way handshake to build a trust relation between the master and the slave node. The appendix's design does not handle threats resulting from an \emph{Insider} threat, including attacks aimed against the important BMC grandmaster election.

We suggest to deprecate the rather complicated security extension, in favor of our proposed PTP-specific defense scheme. Instead of moving towards a more complex symmetric-key cryptographic solution, we propose a public-key-based security solution specially tailored to the PTP protocol. In our solution we take advantage of the already defined \texttt{ANNOUNCE} message along with the \texttt{sequenceId} field, and present an asymmetric cryptographic solution with very light message overhead. Our solution is based on the basic difference between the 2 protocol roles: grandmaster and ordinary slave, a difference yet to be addressed by prior state of the art proposals.

We implemented all our attacks and cryptographic solutions and found that the modified protocol has excellent, practical performance on standard of-the-shelf computers. We believe that these additional security countermeasures will help to provide a complete security extension to the protocol, thus making the IEEE 1588 standard a more secure and robust network solution.


\textbf{Acknowledgements:} We thank D.J. Bernstein and Tanja Lange for suggesting the use of EdDSA and the Ed25519 curve in our implementation.


\newpage

\renewcommand{\theequation}{A-\arabic{equation}}


\setcounter{equation}{0}  

\setcounter{figure}{0}  



\bibliographystyle{plain}

\bibliography{thesis}{}

\end{document}